# The Least Action and the Metric of an Organized System

Georgi Georgiev[1*], Iskren Georgiev [2]

1. Physics and Astronomy Department, TUFTS University, Medford, MA, USA 02155
   Current e-mails: georgi@alumni.tufts.edu; ggeorgie@assumption.edu

2. Physics and Astronomy Department, Sofia University, Sofia, Bulgaria 1000

*To whom all correspondence should be addressed.

# Abstract

In this paper we formulate the Least Action Principle for an Organized System as the minimum of the total sum of the actions of all of the elements. This allows us to see how this most basic law of physics determines the development of the system towards states with less action - organized states. Also we state that the metric tensor can describe the specific state of the constraints of the system, which is its actual organization. With this the organization is defined in two ways: 1. A quantitative: the action I. 2. A qualitative: the metric tensor $g_{\mu\nu}$. These two measures can describe the level of development and the specifics of the organization of a system. We consider closed and open systems.

# Introduction:

Every single process in Nature obeys the Principle of Least Action. For any activity, and in any system, simple or complex, the system spontaneously calculates which path will use least effort for that process.

Our purpose here is very general, to state that all of the structures in the Universe exist, because they are in their state of least action or tend towards it. This state is called organization. If the action is increased by changes in this organization, the system is destroyed, or drops in its level of development as measured by the action.

This is a first of a series of papers that will have as a purpose the description of the process of complexification.

Our goal is to ask more questions and find future directions than to give ready answers. Even the answers of some of the questions cause more questions. This area is very young and in very early stages of development, so almost nothing is worked out. That is why it is very fruitful for future development. We want to leave the feeling that this is an exciting beginning, but not something already established, where there are authorities and canons. There is a full freedom for new discoveries. The theory will change, twist and have unexpected turns. We need to solve different new problems with these two tools. Here are some of the questions:

What form the principle of least action has in very highly developed systems? What new theories and mathematical tools do we need to describe the organization of these systems?

The Principle of Least Action is sufficient to describe simple physical systems. We may need to change it in order to apply it to very highly organized systems. But certainly the laws for the more developed systems will be derived from this principle.

The history of the Principle of Least Action is very long [1-4] and has been very actively applied recently to the widest class of phenomena [5-44]. It is an universal principle, and we still do not know the full extend of its power. Pierre de Maupertuis in *Essai de cosmologie* in 1750 [1] stated the Law of the Least Action as a "universal principle from which all other principles naturally flow". Later Euler, Lagrange, Hamilton, Fermat and many others refined it and applied it to different areas of Physics [45]. Jacobi's form of the Principle refers to the path of the system point in a curvilinear space characterized by the metric tensor. The Hertz's principle of least curvature says that a particle tends to travel along the path with least curvature, if there are not external forces acting on it [45].

Recently there is an enormous amount of literature published on the nonlinear processes in the Universe, complexity, organization and self-organization. This explosion has precipitated in a great quantity of excellent work on the subject, where indispensable accounts for Complexity have been made [46-80]. The approach became a mainstream of our scientific endeavor to understand the world. The view of the unity of all of the processes in the Universe from the Big Bang to the most complex structures now is a standard and mainstream in today's science. We argue that all these processes are rooted in Physics and have physical explanation. People are rushing to mine this new gold of scientific laws waiting to be discovered. New papers, books and meetings at Universities are firing this rapid development. Institutes were formed to study the complex systems:

The New England Complex Systems Institute and Santa Fe Institute in the USA. There are many other institutions in Europe, Asia and Australia and perhaps the seeds are in every other part of the World.

Bertalanffy [66] marked the beginning of a new approach towards the different types of systems. Eric Chaisson [47, 48] defined "Energy Rate Density" as a measure of the organization and complexity of a system, and establishes the term "Cosmic Evolution". Attempts have been made to use the Least action principle to describe certain very limited range of mainly biological systems. Variel and Johnson [17], and Johnson [18] hypothesize that an ecosystem must be structured on interaction between this principle of least action and an overriding Principle of Most Action. Sergeev et. al. [20] suggest to use the Hamilton-Lagrange principle of least action as a descriptive tool for qualitative measure of self-organization of functional biotechnical systems. The Least-Action Principle is applied even in computer science by Grard A. Langlet [52,53].

As a contribution to this process we want to suggest the use of two very powerful techniques, not yet fully utilized in this new search to describe universally the nonlinear, non-equilibrated, organized and complex systems, for the whole sequence of structures from the Big Bang, to the highly sophisticated structures in our times. These tools are universal and can be applied to the simplest and most complex systems, allowing us to escape from the limitations of the tools that allow us to describe just one portion of the whole process of development from the pure energy after the beginning of the Universe, to the today's most complex systems. These are: 1. A quantitative: the action I, and 2. A qualitative: the metric tensor $g_{\mu\nu}$. In the future, there may be additional approaches added, but nevertheless the power of these two will not be undermined.

The Principle of Least Action is the basis of Physics. For the most of it's history it has been applied mainly to a very limited class of simple systems. The laws of motion for a particle can be derived from the least action principle, that shows that the particle is moving along a geodesic [45]. Important consequences can be derived, if it is applied to a more complex system. Then, the situation is quite different, because now all of the particles have to achieve in the same space and time, a state with least action. But in a system, they cannot move along their geodesics because there are constraints. Then their path becomes curved, and this curvature can be described by the metric tensor $g_{\mu\nu}$ [82].

$$ds^2 = \sum_{\mu\nu} g_{\mu\nu}\, dx^\mu dx^\nu \qquad (1)$$

In an organized system, the action of a single element will not be at minimum, because of the constraints, but the sum of all actions in the whole system will be at minimum. The action of a single element is not maximum as well, because by definition this will destroy the system, so this intermediate state represents an optimum. The reorganization of the system to achieve it is a process of optimization. All of the elements find their path of least possible action in the curved by the constraints phase space. The geodesic is modified each time the position of any of the elements or the constraints of the system is changed.

The elements apply work on the constraints to modify the organization and minimize the action, which takes finite amount of time, making the reorganization a process. The state of the constraints in the system that determines the sum of the actions of its elements is called organization. So the process of self-organization of the systems can be called a "Process of achieving a least action state by a system". It could last billions of years or indefinitely. The laws of achieving this least action state will be the laws of development of the system, with the least action state as a final state (if new elements are added all of the time, making the system open, then the least action state can never be reached and the process is infinite). We can consider two cases: closed systems, where the number of elements and constraints is fixed, and the least action state can be reached in finite time, and open systems, where there is a constant flow and change of the number of elements, constraints and the energy of the system, and in this case the system will always approach this state, which is an attractor for the system, but will never reach it. It will be in a constant process of reorganization.

This process acts in the whole Universe, and can be called a "Law of Progressive Development of the Universe".

Here we show how a systemic view can describe the state of minimum action for the entire system, which is not the state of minimum action of an individual element. Then the action of the whole system is defined as the sum of the actions of all of the elements, and the action of each of them must be such, that it ensures least action for the sum of the actions of all elements in the system. This minimum action state of the system is realized through reconfiguration of its constraints. The Gauss Principle of Minimum Constraint [81], can be used to derive the minimum constraint configuration of the system. This is actually the process of progressive development of the system reconfiguring the constraints, so they ensure a state of minimum sum of actions for the elements.

# Two cases:

### I. Closed systems:

The principle for least action states, that the actual motion of a conservative dynamical system between two points, occurs in such a manner, that the action has a minimum value in respect to all other paths between the points, which correspond to the same energy.

The classical definition of the principle of least action is [45]:

$$\Delta \int_{t_1}^{t_2} p_i q_i dt = 0 \qquad (2)$$

the variation of the path is zero for natural process, or the nature acts in the simplest way.

The action:

$$I = \int_{t_1}^{t_2} L dt \qquad (3)$$

Where L is the lagrangian of the system:

$$L = T - V \qquad (4)$$

Here T and V are the kinetic and the potential energy of the system respectively.

Action has the dimensions of energy multiplied time (in joule-seconds).

For the motion of the system between time $t_1$ and $t_2$, the lagrangian L has a stationary value for the correct path of motion.

This can be summarized as the Hamilton's Principle.

$$\delta I = \delta \int_{t_1}^{t_2} L\, dt = 0 \qquad (5)$$

If we extend this to a system of two elements, then the above principle (2) will become: the variation of the sum of the actions of the individual elements in the system, must be zero.

If we have a single element and one constraint, according Least Action Principle, the element tends to move on a straight line. But if there is a constraint, the action of the element will increase by circumventing it. In order to minimize the constraint and therefore its action, the element has to apply force on the constraint. This is the basic model that we consider. It can be complicated by adding more and more elements to the system, and finding the least action and least constraint state. We consider the sum of the actions for a system.

For a system with two elements:

$$\delta(I_1 + I_2) = 0 \qquad (6)$$

For a larger system the variation of the sum of the actions of all of the elements tends toward zero.

$$\delta\left(\sum_i I_i\right) = 0 \qquad (7)$$

Or if we expand:

$$\delta\left(\sum_i I_i\right) = \delta\left(\sum_i \int_{t_1}^{t_2} L_i\, dt\right) = \delta\left(\sum_i \int_{t_1}^{t_2} (T_i - V_i)\, dt\right) = 0 \qquad (8)$$

This formulation says that the least action for a system is the least sum of the actions of all elements in the system. As mentioned above, the least constraint state is achieved through the work done by the elements on the constraints, reconfiguring them, tending toward a state described by equation (8). The elements exercise force on the constraints, which is proportional to their action, changing them to state with less action by doing work - organizing them. The elements themselves act as constraints for each other. This makes the problem even more nonlinear and complicated. Here can be used the principle of least constraint by Gauss in *Über ein neues allgemeines Grundgesetz der Mechanik* (1831).

Let's consider an example. The simplest possible situation is one element and one constraint. Then if the constraint is obstructing the motion of the element, its action will be increased. But the element will apply force to turn the constraint in such a way, that the next motion will occur with less action, or if the constraint is completely out of the way, with least action. This implies a periodic motion of the elements in an organized system.

Now lets consider two elements and one constraint. If the action of one of them is minimum, when the configuration of the constraint is such that the action for the other is maximum, then the action for the whole system is maximum, and it is not going to be preserved. The action of each of the elements must be such that each of them has some minimal action, but not such that one of them has least action, and the other has an infinite action. Then the sum of the two actions will be infinity, and the system is impossible to exist.

For system with a very large number of elements, the constraints have to be positioned in such a way, that they will ensure the least sum of all individual actions of the elements. Then, similarly to the way we derive the laws of motion, we can derive the laws of progressive development of the whole organized system of any type and any level of development. The advance of the organization in the process of progressive development of a system then is a search for minimizing the action of the system as a whole, by the individual elements searching for minimum action.

**II. Open systems.**

Adding and subtracting elements and energy to and from a system and recalculating its least action state by expression (8) is what is making it open. For an open system there is a constant flow of matter and energy. It means that the system has to constantly reevaluate its least action state, or that it works toward being in least action state, but because of the constant changes, this state is never reached. The system only tends towards this state, it is driven by this tendency, and reorganizes further. This system is not deterministic anymore. We cannot predict the positions of all of the elements in the system in the future.

A particle achieves least action state, by moving along geodesic ( straight line in linear space). When several particles and constraints are added to the first one, the state can be achieved in some longer time. But if the number of particles and constraints becomes very large n>>1 , and when the system is open, the end state could never be fixed. The individual geodesic will be a constantly changing trajectory in curved phase space of the system, described by the metric tensor $g_{\mu\nu}$, even when the element is in the middle of the motion.

So the organization is defined by the action. Decrease in action, corresponds to increase in the in the organization.

We can summarize the following definitions:

- Organization: the state of the constraints that determines the action and metric of a system.

- To organize: to change the state of the constraints in order to ensure less action.

- For a closed system the final least action state of the system can be achieved.

- For an open system the attractor is the least action state but because of the flow of matter and energy it can only be approached and never reached. One open system is better organized if it has less action than another open system with the same number of elements.

## Conclusions:

Formulation of the Principle of Least Action for an organized system has been done. In this paper we suggest a new way of characterization of organized physical systems. We propose two methods of description of organization in closed and open systems.

1. A quantitative method: the action I.

2. A qualitative method: the metric tensor $g_{\mu\nu}$.

The quantitative is a scalar, and measures the absolute change in the organization of the system. The qualitative is a tensor, describing the specifics the organization of the system. One level of organization with given action I, can be realized through infinite number of combinations of constraints described by the metric tensor $g_{\mu\nu}$.

We extended the least action principle for a whole organized system, to derive it as a zero variation of the sum of actions of all elements. This leads to important consequences, because the minimum of the sum of all actions is not necessarily the minimum of the action of each element, but some optimum value. It is a minimum for the given constraints and configuration of elements. When new elements are introduced, new configurations and new minimum state is reached. For open systems a final state is never achieved, because of the constant recalculation of the least action state, due to the constant flux of energy and matter.

The most developed system has organization in a state that exerts least constraint, and by that ensures least action of the system. We argue that the action of the whole system determines the level of organization and can be used as a quantitative measure of degree of progressive development for all systems in the Universe, starting from the Big Bang and ending with the technological society. This can provide framework for study, classification and comparison between all systems and the processes leading to the increase in organization. It is applicable for all systems physical, chemical, biological, economical and technological. We can use this understanding for optimizing systems in our society making it more efficient and highly organized.

**Literature:**

1. Pierre de Maupertuis, *Essai de cosmologie*, (1750).

2. D.I. Youssouf, A. Guran "The principle of least action: history of a long controversy," *Transactions of the Canadian society for mechanical eng.* **24,** (1b) 285 (2000).

3. M. Stoltzner "*Vienna indeterminism: Mach, Boltzmann, Exner,*" *Synthese* **119,** (1-2) 85 (1999).

4. P. Dias. "Euler's "harmony" between the principles of "rest" and "least Action"- the conceptual making of analytical mechanics," *Archive for history of exact sciences* **54,** (1) 67 (1999).

5. V.V. Kozlov, "*Principle of least action and periodic-solutions in problems of classical mechanics*," *J Appl Math Mec* **40**, (3) 363 (1976).

6. S. Gavrilyuk, H. Gouin "A new form of governing equations of fluids arising from hamilton's principle," *International journal of engineering science* **37,** (12) 1495 (1999).

7. G.F. Filippov, S.V. Mokhov, A.M. Sytcheva, K. Kato, S.V. Korennov "Analysis of equations of antisymmetrized molecular dynamics for some simple systems," *Physics of atomic nuclei* **62,** (1) 95 (1999).

8. I.P. Guk **"**Lagrange formalism for particles moving in a space of fractal Dimension," *Technical physics* **43,** (4) 353 (1998).

9. Y.J. Xie, C.H. Liang, "Generalized principle of least action in electromagnetism and its Applications," *Chinese science bulletin* **43,** (9) 732 (1998).

10. H.H. Diebner, O.E. Rossler, "A deterministic entropy based on the instantaneous

phase space volume," *Zeitschrift fur naturforschung section A-A. Journal of physical sciences* **53,** (1-2) 51 (1998).

11. W.G. Hoover, "Time reversibility in nonequilibrium thermomechanics," *Physica D* **112,** (1-2) 225 (1998).

12. G. Cavalleri, E. Tonni, "Negative masses, even if isolated, imply self-acceleration, hence a catastrophic world," *Nuovo cimento della societa italiana di fisica b-general physics relativity astronomy and mathematical physics and methods* **112,** (6) 897 (1997).

13. K. Fujikawa, "Path integral of the hydrogen atom, jacobi's principle of least action and one-dimensional quantum gravity," *Nuclear physics B* **484,** (1-2) 495 (1997).

14. A. Granik, H.J. Caulfield, "Fuzziness in quantum mechanics," *Physics essays* **9,** (3) 496 (1996).

15. X.Y. Yu, W.C. Henneberger, "Path-dependent phase shifts in isolated systems," *International journal of theoretical physics* **35**, (2) 333 (1996).

16. M. Lauster, "Conservation laws in non-linear dissipative systems," *International journal of non-linear mechanics* **30,** (6) 915 (1995).

17. P. Vanriel, L. Johnson. "Action principles as determinants of ecosystem structure - the autonomous lake as a reference system," *Ecology* **76,** (6) 1741 (1995).

18. L. Johnson, "Pattern and process in ecological-systems - a step in the development of a general ecological theory," *Canadian journal of fisheries and aquatic sciences* **51,** (1) 226 (1994).

19. V.E. Tarasov "Quantum dissipative systems .1. Canonical quantization and quantum liouville equation," *Theoretical and mathematical physics* **100,** (3) 1100 (1994).

20. E.V. Sergeev, E.F. Aksyuta, O.Y. Boxer, "The hamilton-lagrange method as a theoretical foundation for study of self-organization of model and realistic biotechnical system," *Biofizika* **40,** (1) 132 (1995).

21. V.A. Vladimirov, H.K. Moffatt, "On general transformations and variational-principles for the magnetohydrodynamics of ideal fluids .1. Fundamental principles," *Journal of fluid mechanics* **283,** 125 (1995).

22. M. Susperregi, J. Binney, "The principle of least action and clustering in cosmology," *Monthly notices of the royal astronomical society* **271,** (3) 719 (1994).

23. B. Stephanis, B.K. Papadopoulos, N. Elias, "Investigation of multivariable systems," *Applied mathematical modelling* **18,** (11) 628 (1994).

24. D. Pollard, J. Dunningdavies, "A consideration of the possibility of negative mass," *Nuovo cimento della societa italiana di fisica b-general physics relativity astronomy and mathematical physics and methods* **110,** (7) 857 (1995).

25. V.B. Magalinsky, M. Hayashi, H.V. Mendoza, "Variational approximation of density-matrix on the basis of ground-state - thermalization of vacuum," *Journal of the physical society of Japan* **63,** (8) 2930 (1994).

26. F. Cooper, H. Shepard, P. Sodano, "Solitary waves in a class of generalized korteweg-devries equations," *Physical Review E* **48,** (5) 4027 (1993).

27. S.V. Shabanov, "Quantum and classical mechanics of q-deformed systems," *Journal of physics a-mathematical and general* **26,** (11) 2583 (1993).

28. Y. Sobouti, S. Nasiri, "A phase-space formulation of quantum state functions," *International journal of modern physics B* **7,** (18) 3255 (1993).

29. F. Cooper, C. Lucheroni, H. Shepard, P. Sodano, "Variational method for studying

solitons in the korteweg-devries equation," *Physics letters A* **173,** (1) 33 (1993).

30. S. Peterson, "The geometry of virtual work dynamics in screw space," *Journal of applied mechanics-transactions of the ASME* **59,** (2) 411 (1992).

31. T. Chan, "Indefinite quadratic functionals of gaussian-processes and least-action paths," *Annales de l institut henri poincare-probabilites et statistiques* **27,** (2) 239-271 (1991).

32. J. Mhalla, "A new derivation of classical diffusion-equations from the principle of least action," *Journal de chimie physique et de physico-chimie biologique* **88,** (1) 1 (1991).

33. I.Y. Krivskii, V.M. Simulik, "Noether analysis of zilch conservation-laws and their generalization for the electromagnetic-field .1. Use of different formulations of the principle of least action" *Theoretical and Mathematical Physics* **80,** (2) 864 (1989).

34. I.Y. Krivskii, V.M. Simulik, "Noether analysis of zilch conservation-laws and their generalization for the electromagnetic-field .2. Use of poincare-invariant formulation of the principle of least action," *Theoretical and Mathematical Physics* **80,** (3) 912 (1989).

35. G. Dickel, "Hamilton principle of least action in nervous excitation," *Journal of the chemical society-faraday transactions I* **85,** 1463, part 6 (1989).

36. T. Itoh, K. Abe, "Discrete lagrange equations and canonical equations based on the principle of least action," *Applied mathematics and computation* **29**, (2) 161 (1989).

37. J.G. Papastavridis, "Rayleigh principle via least action," *Journal of sound and vibration* **113**, (2) 395 (1987).

38. J.G. Papastavridis, G. Chen, "The principle of least action in nonlinear and-or

nonconservative oscillations," *J Sound Vib.* **109**, (2) 225 (1986).

39. J.G. Papastavridis, "The principle of least action as a lagrange variational problem - stationarity and extremality conditions," *Int J Eng Sci* **24**, (8) 1437 (1986).

40. J.C. Zambrini, "Maupertuis principle of least action in stochastic calculus of variations," *J math phys* **25**, (5) 1314 (1984).

41. R. Broucke, "On the principle of least action - some possible generalizations," *Hadronic J.* **5**, (5) 1901 (1982).

42. H. Nadj, J. Nadj, M. Nadj, "Relativity as consequence of the principle of least action," *Am phys soc* **26**, (3) 472 (1981).

43. A. Peton, "Weyl-theory and principle of least action - application to study of motions," *Astrophys space sci* **69**, (1) 147 (1980).

44. A.I. Perrote, A.N. Yavriyan, "Principle of least action in reliability," *Theory Eng Cybern* **16**, (6) 101 (1978).

45. H. Goldstein, *Classical Mechanics*, (2nd ed., Addison Wesley, 1980).

46. Y. Bar-Yam, *Dynamics of Complex Systems*, (Addison Wesley, 1997).

47. E. J. Chaisson, *The cosmic Evolution*, (Harvard, 2001).

48. E. J. Chaisson, "The cosmic environment for the growth of complexity," *BioSystems,* **46,** 13 (1998).

49. S. J. Goerner, *Chaos and the evolving ecological Universe*, (Gordon and Breach, 1994).

50. S. N. Salthe, *Development and Evolution,* (MIT Press, 1993).

51. S. Goontilake, *Evolution of Information*, (Pinter Pub Ltd, 1991)

52. G. A. Langlet, "Towards the Ultimate APL-TOE, APL" *Quote-Quad,* **22**, 1 (APL92,

St Petersburg, Russia, July 1992) p. 118-132.

53. G. A. Langlet, "The Power of Boolean Computing in APL," *SEAS 94* (B-La Hulpe, IBM Conf. Centre, April 1994).

54. R.B. Daniel, O.W. Edward, *Evolution as entropy*, (University of Chicago Press, 1988).

55. I. Stengers, I. Prigogine, *The End of Certainty*, (Free Press, 1997).

56. V.F. Turchin, *The Phenomenon of Science: A cybernetic approach to human evolution*, (Columbia University Press, 1977).

57. P. Coveney, R. Highfield, *Frontiers of Complexity*: *The search for order in a chaotic world*, (Fawcett Columbine, 1995).

58. J. Casti, *Would-be Worlds*, (Wiley, 1997).

59. J. Holland, *Emergence: from Chaos to Order*, (Addison Wesley, 1998).

60. E. Jantsch, *The Self-Organizing Universe: Scientific and human implications of the emerging paradigm of evolution*, (Pergamon Press, 1980).

61. S. Kauffman, *The Origins of Order*: *Self-organization and selection in evolution,* (Oxford, 1993).

62. R. Lewin, *Complexity*: *life at the edge of chaos*, (Macmillan, 1992).

63. P. Bak, *How Nature Works: The science of self-organized criticality*, (Copernicus, Springer Verlag, 1996).

64. M. Waldrop, *Complexity: The emerging science at the edge of order and chaos*, (Simon and Schuster, 1992).

65. W. Zurek (ed.), *Complexity, Entropy and the Physics of Information*, (Addison Wesley, 1989).

66. L. von Bertalanffy, *General Systems theory,* (Brasillier, 1968).

67. H. Haken, *Synergetics*, (Springer-Verlag, 1978).

68. M. Eigen, P. Schulster, *The hypercycle - a principle of natural selforganization*, (Springer, 1979) .

69. E. Jantsch (ed.), *The evolutionary vision*, (Westview press, 1981).

70. E. Jantsch, *The Self-Organizing Universe*, (Pergamon, Oxford, 1980).

71. Kauffman S., *Investigations*, (Oxford, 2000).

72. E. Laszlo, *Evolution: the grand synthesis*, (Shambhala, 1987).

73. J. Lovelock, *The ages of Gaia*, (Oxford, 1988).

74. J. S. Nicolis, *Dynamic of hierarchical systems*, (Springer-Verlag, 1986).

75. I. Prigogine, *Introduction to the Thermodynamics of Irreversible Processes*, (Wiley, 1961).

76. G. Nicolis, I. Prigogine, *Exploring Complexity*, (W H Freeman & Co., 1989).

77. E. O. Wilson, *Consilience: The Unity of Knowledge*, (Knopf, 1988).

78. F. Capra, *The web of life: a new scientific understanding of living systems*, (Anchor books, 1996).

79. R. U. Ayres, *Information, Entropy, and Progress: A New Evolutionary Paradigm* (Springer Verlag, 1994).

80. J. D. Collier, C. A. Hooker, "Complexly Organised Dynamical Systems," *Open Systems & InformationDynamics* **6**, 241 (1999).

81. Gauss, *Über ein neues allgemeines Grundgesetz der Mechanik*, (1831).

82. G.B.Arfken, H.J. Weber, *Mathematical Methods for Physicists*, (4th ed., Academic Press, 1995).